\documentclass[aps,prl,reprint,superscriptaddress]{revtex4-1}

\usepackage{graphicx}
\usepackage{dcolumn}
\usepackage{bm}

\begin{document}

\title{Lifting of the Landau level degeneracy in graphene devices in a tilted magnetic field}

\author{F.~Chiappini}
\email[]{f.chiappini@science.ru.nl}
\affiliation{High Field Magnet Laboratory (HFML-EMFL) and Institute for Molecules and Materials, Radboud University, Toernooiveld 7, 6525 ED Nijmegen, The Netherlands} 

\author{S.~Wiedmann}
\affiliation{High Field Magnet Laboratory (HFML-EMFL) and Institute for Molecules and Materials, Radboud University, Toernooiveld 7, 6525 ED Nijmegen, The Netherlands}

\author{K.~Novoselov}
\affiliation{Department of Physics, University of Manchester, M13 9PL Manchester, United Kingdom}

\author{A.~Mishchenko}
\affiliation{Department of Physics, University of Manchester, M13 9PL Manchester, United Kingdom}

\author{A.~K.~Geim}
\affiliation{Department of Physics, University of Manchester, M13 9PL Manchester, United Kingdom}

\author{J.~C.~Maan}
\affiliation{High Field Magnet Laboratory (HFML-EMFL) and Institute for Molecules and Materials, Radboud University, Toernooiveld 7, 6525 ED Nijmegen, The Netherlands}

\author{U.~Zeitler}
\email[]{u.zeitler@science.ru.nl}
\affiliation{High Field Magnet Laboratory (HFML-EMFL) and Institute for Molecules and Materials, Radboud University, Toernooiveld 7, 6525 ED Nijmegen, The Netherlands}

\begin{abstract}
We report on transport and capacitance measurements of graphene devices in magnetic fields up to 30~T. In both techniques, we observe the full splitting of Landau levels and we employ tilted field experiments to address the origin of the observed broken symmetry states. In the lowest energy level, the spin degeneracy is removed at filling factors $\nu=\pm1$ and we observe an enhanced energy gap. In the higher levels, the valley degeneracy is removed at odd filling factors while  spin polarized states are formed at even $\nu$. Although the observation of odd filling factors in the higher levels points towards the spontaneous origin of the splitting, we find that the main contribution to the gap at $\nu= -4,-8$, and $-12$ is due to the Zeeman energy.
\end{abstract}


\maketitle

One of the prominent consequences of the Dirac-like nature of charge carriers in graphene is the half-integer quantum Hall effect. The Hall conductance is quantized to half-integer multiples of $4e^2/h$, reflecting the spin and valley degeneracy of the Landau levels (LLs) at filling factors $\nu=4(N+1/2)=\pm 2, \pm 6, \pm 10,$... \cite{NovoselovNat2005, ZhangNat2005}, where $N$ is the LL index. 

However,  electron-electron interactions and explicit symmetry breaking fields, such as the Zeeman splitting, can lift the LL degeneracy, leading to the observation of the integer quantum Hall effect in intermediate states. The origin and the possible spin and/or valley polarization of the  broken symmetry states  has been the subject of considerable theoretical interest \cite{AliceaPRB2006, ShengPRL2007, GoerbigRevModPhys2011, HerbutPRB2007, GusyninPRB2006, LuoPRB2013, RoyPRB2011, ShengPRL2007} and experimental investigations \cite{ZhangPRL2006, JiangPRL2007, ZhangPRB2009, GiesbersPRB2009, CheckelskyPRL2008, YoungNatPhys2012,ZhaoPRL2012, AmetPRL2014}.

Earlier experimental works on graphene on SiO$_2$ reported the partial splitting of $N=\pm 1$ at $\nu=\pm 4$ \cite{ZhangPRL2006, JiangPRL2007, ZhaoPRL2012} as a single particle effect due to the Zeeman energy, and the full splitting in the lowest LL driven by electron-electron interactions \cite{JiangPRL2007}. Experiments on cleaner devices deposited on hexagonal boron nitride (hBN) showed the full sequence of integer filling factors and addressed the role of electron-electron interactions in the LLs splitting \cite{YoungNatPhys2012, YuPNAS2013}, revealing enhanced gaps in the higher LLs and skyrmion mediated transport in $N=-1$ \cite{YoungNatPhys2012}. These observation have been attributed to the quantum Hall ferromagnetism (QHF) in graphene \cite{NomuraPRL2006}.
 
In this Rapid Communication we report on the LL splitting in graphene encapsulated between two layers of h-BN and we investigate the nature of the states occurring at integer $\nu$ due to the lifting of the LL degeneracy. Thermally activated transport in both perpendicular and tilted magnetic fields up to 30 T, supported qualitatively by capacitance spectroscopy \cite{YoungNature2013}, enables us to probe the origin of the states at $\nu=-1,-3,-4, -7, -8, -12$.
We show that in the lowest LL, a spin unpolarized state is formed at $\nu=0$ and the spin degeneracy is removed at $\nu=\pm 1$. In the higher levels, the even $\nu$ separate two fully spin polarized states while the odd $\nu$ originate from the lifting of the valley degeneracy, supporting the findings of Ref.~\onlinecite{YoungNatPhys2012}. We find that, for $|N|\geq 1$, the gap at half filling is set by the Zeeman energy and we demonstrate that the interactions, although relevant in determining the splitting of the higher levels, are not the dominant energy scale for $|N|>0$ in our devices.

\begin{figure}
 \includegraphics{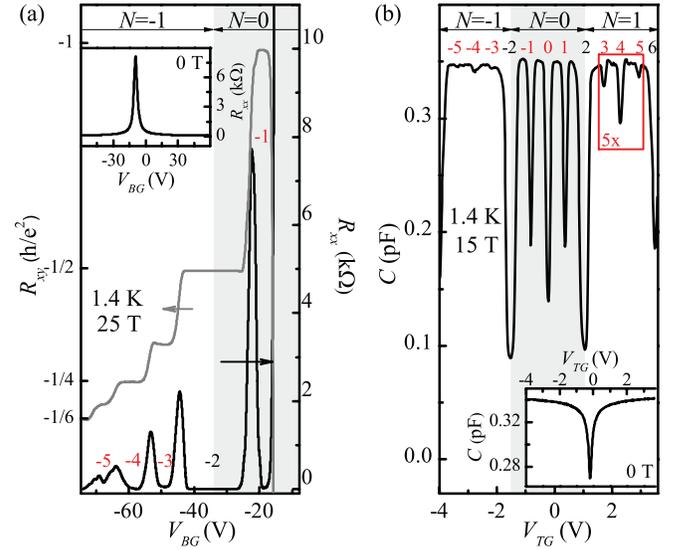}%
 \caption{\label{fig:fig1}(Color online) Transport and capacitance measurements at 1.4~K. (a) $R_{xx}$ (black line) and $R_{xy}$ (gray line) as a function of the back gate voltage $V_{BG}$ at 25~T. Inset: $R_{xx}$ as a function of $V_{BG}$ at 0 T. (b) $C$ as a function of the top gate voltage $V_{TG}$ at 15~T. The curve inside the red box is expanded five times and shifted in order to match the constant background of the original curve.  The numbers close to the minima of $R_{xx}$ and $C$ indicate the filling factors. Inset: $C$ as a function of $V_{TG}$ at 0~T.}
 \end{figure}
 
We focus on two single layer graphene devices. In both devices the graphene flake is sandwiched between two h-BN flakes. Device A is a Hall bar ($W\approx$~1.3 $\mu$m, $L/W=$~2) for standard magneto-transport measurements. The longitudinal ($R_{xx}$) and Hall ($R_{xy}$) resistances are measured as a function of the back gate voltage ($V_{BG}$) using a low noise lock-in technique with a 10~nA excitation current at 13~Hz. Measurements at the charge neutrality point (CNP) in a magnetic field are performed in a constant voltage configuration with a 100 $\mu$V excitation. Device B is a graphene-hBN-Au capacitor, similar to the ones described in Ref.~\onlinecite{YuPNAS2013}, with a 37~nm thick h-BN flake between the graphene and the Au electrode. The capacitance $C$ is measured as a function of the DC voltage ($V_{TG}$) applied between the top gate and the graphene sheet using a capacitance bridge (AH2700) with 30~mV AC excitation at 20~kHz. $C$ embodies two major contributions: the geometrical capacitance $C_G=$~0.346~pF and the quantum capacitance $C_Q$, which is directly proportional to the density of states (DOS)
\cite{LuryiApplPhysLett1988} of graphene. Both devices were placed in a variable temperature $^{4}$He cryostat in a Bitter magnet on a sample holder which allows \textit{in situ} rotation.

 We first characterize our devices at $T$=1.4~K in the absence of a magnetic field. As illustrated in the insets of Figs.~\ref{fig:fig1}(a) and 1(b) both samples are $n$-doped and the CNP is situated at $V_{BG}^{CNP}\approx-10$~V for device A and $V_{TG}^{CNP}\approx -0.2$~ V for device B, corresponding to a residual electron concentration $n_A \approx 6 \times 10^{11}$~cm$^{-2}$ and $n_B  \approx 1 \times 10^{11}$~cm$^{-2}$, respectively. The field effect mobility has been extracted for device A according to Ref.~\onlinecite{DeanNatNanotech2010} and is found to be $\mu=4\cdot 10^{4}$ cm$^2$/Vs.

In a magnetic field, both samples show the full lifting of the Landau level degeneracy. In Fig. \ref{fig:fig1}(a) we show $R_{xx}$ of device A at 25 T and 1.4 K for the hole side ($V_{BG} <V_{BG}^{CNP}$). Minima in $R_{xx}$ and quantized Hall plateaus in $R_{xy}$ at $\nu= -1, -3, -4$, and $-5$  are well developed and more pronounced compared to those observed in the electron side. We will therefore focus our further analysis on the LLs for the holes. For device B [Fig.~\ref{fig:fig1}(b)], clear minima in the capacitance measurements are detected at each integer value of $\nu$ in $N=0$ and $N=\pm 1$  at 15 T and 1.4 K. It is worth noting that the lifting of the LL degeneracy evolves progressively with the magnetic field, first in the lowest LL and then in the higher levels and, within each level, first at half filling and then at quarter filling. Hence, the first filling factor observed is $\nu=0$, appearing as a clear minimum in $C$ for $B\geq 5$~T and as a diverging $R_{xx}$ for $B>2$~T [see Fig.\ref{fig:fig4}(c), points connected by the grey line]. Filling factors $\nu=\pm$1 are well developed already at 10 T for both samples. For $B\geq$10~T the full splitting 
of $N=\pm$1 starts to be resolved in the capacitance spectroscopy. In transport measurements, plateaus in $R_{xy}$ appear in the $N=-1$ LL, first at $\nu=-4$ (10~T), then at $\nu=-3$ (12.5~T, and finally at $\nu=-5$ (17.5~T). 

\begin{figure}
 \includegraphics{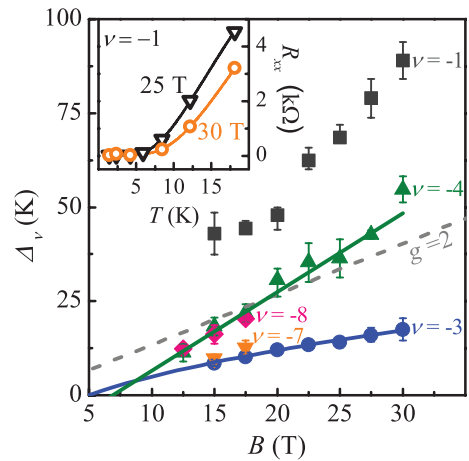}%
 \caption{\label{fig:fig2}(Color online) Activation gaps $\Delta_{\nu}$ as a function of $B$ for $\nu=-1, -3, -4, -7$ and $-8$. The gray dashed line indicates $E_Z$ calculated with $g=2$, and the solid lines  are the linear (green) and square root (blue) fit to the data. Inset: $R_{xx}$ minima of $\nu=-1$ as a function of temperature at 25~T (black triangles) and 30~T (orange circles), and the solid lines are a fit to the experimental data according to the Fermi-Dirac distribution. The relatively large error bars for gaps exceeding 20 K are due to the fact that the temperature range used ($1.4-18$~K) was not large enough to access them more accurately.}
 \end{figure}

Let us now address the sizes of the energy gaps $\Delta_{\nu}$ that are associated with 
the broken symmetry states in a purely perpendicular magnetic field 
(Fig.~\ref{fig:fig2}). We extract  $\Delta_{\nu}$  from temperature-activated transport experiments on device A between 1.4 and 18 K, fitting the experimental data according to the Fermi-Dirac distribution 
$R_{xx}\propto 1/(e^{-\Delta_{\nu}/2k_BT}+1)$, since the size of the gap is comparable 
to $k_{B}T$ in the temperature range under study \cite{KurganovaSolidStateComm_2010}. The inset of Fig.~\ref{fig:fig2} shows the minima of $R_{xx}$ as a function of temperature for $\nu=-1$ at 25 (triangles) and 30 T (circles) and the fits (solid lines) as an example of typical fitting traces. Since there is some uncertainty in the range of applicability of the activated transport assumption, at some magnetic field values we had to perform different fits considering each time a different temperature range. The values of $\Delta_\nu$ plotted in Fig.~\ref{fig:fig2} are an average over the values obtained fitting in the different temperature ranges, and the error bars represent the spread of all the values obtained due to the different fits.

As can be seen in Fig.~\ref{fig:fig2}, the size of the energy gaps increases with the magnetic field for each filling factor. In the higher LLs ($N=-1$ and $-2$) we can distinguish between two different energy scales, one for the gaps at even $\nu$ and one for the gaps at odd $\nu$. 
 The size of $\Delta_{-4}$ and $\Delta_{-8}$ increases linearly with the magnetic field and the gaps for both filling factors fall on the same line. The gap size can be described by $\Delta=g\mu_BB-\Gamma$; the first term is the Zeeman energy ($E_Z$) and $\Gamma$ is the Landau level broadening. A fit to the experimental data yields $g=~3.1\pm 0.1$ and $\Gamma=~14.4\pm 1.5$ K. The enhancement of $g$ compared to its bare value 2 is probably due to exchange interaction \cite{KurganovaPRB_2011}.
 
The gaps at odd filling factors within the $N=-1$ and $N=-2$ LLs, $\Delta_{-3}$ and  $\Delta_{-7}$, are comparable  within the error bars. They are considerably smaller than $E_Z$ and their field dependence can be fitted by a square root function with a finite offset representing a Landau level broadening of  $12\pm 1$~K (blue solid line for of $\Delta_{-3}$). A linear function, which would also reasonably fit the data, leads to a meaningless negative value for $\Gamma$. 

 Though a shallow minimum develops in $R_{xx}$ at $\nu=-5$ for $B> 17.5$ T, we are not able to extract the activated gap in the considered temperature range  since the $R_{xx}$ minimum is visible only at the lowest temperatures. Thus, we conclude that the state at $\nu=-5$ is weaker than the state at $\nu=-3$, as also suggested by the capacitance signal where dips at $\nu=\pm5$ are not as pronounced as the ones at $\nu=\pm3$ [see the curve inside the red box in Fig.~\ref{fig:fig1}(b)].

In addition, Fig.~\ref{fig:fig2} highlights already the different behavior of the states within the lowest LL compared to the ones in the higher levels. Indeed we notice that $\nu=-1$ has the largest energy gap and, in particular, it is much larger  than the size of the gaps of the other odd filling factor $\nu=-3 $ and $\nu=-7$.

\begin{figure}
 \includegraphics{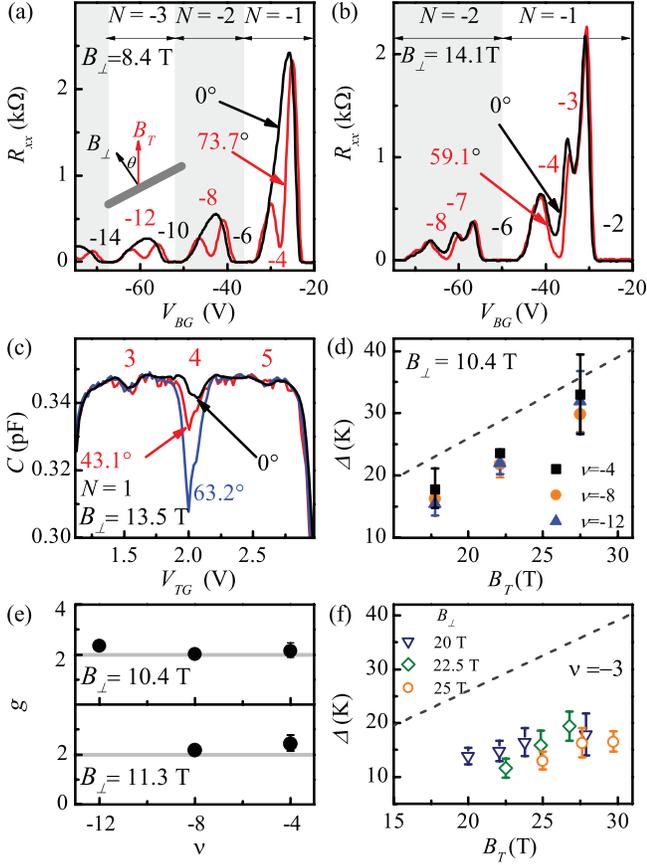}%
 \caption{\label{fig:fig3}(Color online) Splitting of the higher Landau levels in a tilted magnetic field at 1.4 K. (a) $R_{xx}$ as a function of the back gate voltage at $B_\perp=8.4$ T and $\theta=0^\circ$ (black line) and at $\theta=73.7^\circ$ (red line). Inset: Sketch of the tilting configuration. (b) $R_{xx}$ for $N= -1$ and $-2$  at $B_\perp= 14.1$~T and $\theta=0^\circ$  (black line) and $\theta=59.1^\circ$ (red line). (c) Splitting of $N=1$ observed in the capacitance signal at $B_\perp=$ 13.5 T and $\theta=0^\circ$ (black line), $\theta=43.1^\circ$ (red line) and $\theta=63.2^\circ$ (blue line). (d) Activation gaps for $\nu=-4$ (black squares), $\nu=-8$ (orange circles) and $\nu=-12$ (blue triangles) as a function of $B_T$ at $B_\perp=10.4$~T. They gray dashed line represents $E_Z$ for $g=$2. (e) $g$ as a function of $\nu$ at $B_\perp=$ 10.4~T (top panel) and 11.3~T (bottom panel). (f) $\Delta_{-3}$ as a function of $B_T$ at $B_\perp=$20~T (blue triangles), 22.5~T (green squares) and 25~T (orange circles).}
 \end{figure}

\begin{figure}
\includegraphics{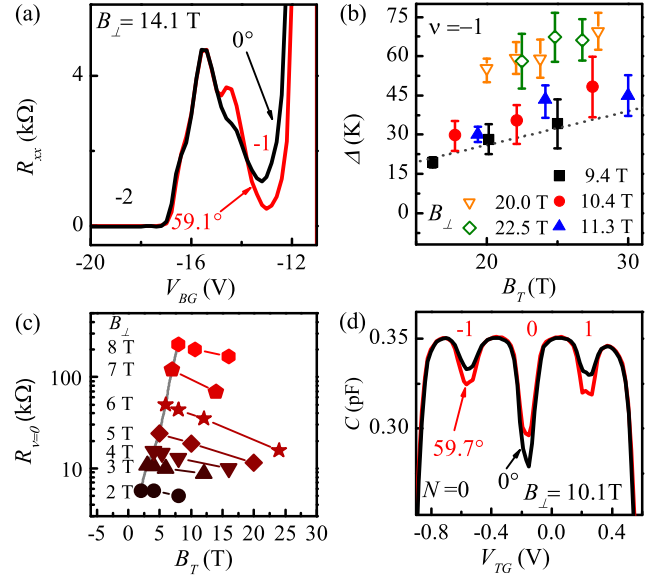}%
\caption{\label{fig:fig4}(Color online) Splitting of $N=0$ in a tilted magnetic field. (a) $R_{xx}$ at $\nu=-1$ in $B_\perp=14.1$~T and $\theta=0^\circ$ (black line) and $\theta=59.1^\circ$ (red line). (b) Activation gap for $\nu=-1$ as a function of $B_T$ at different $B_\perp$; the gray dotted line is $E_Z$ calculated with $g=2$. (c) Value of the resistance maximum at $\nu=0$ in a tilted magnetic field as a function of $B_T$ at different $9.4\leq B_\perp\leq 22.5$~T and 4.2 K. (d) $C$ at $B_\perp=10.1$~T and $\theta=0^\circ$ (black line) and $\theta=59.7^\circ$ (red line).}
\end{figure}

In order to further investigate the origin of the broken symmetry states we perform  magnetotransport and capacitance experiments in a tilted magnetic field. We tilt the sample with respect to the direction of the magnetic field by an angle $\theta$ [see inset of Fig.\ref{fig:fig3}(a)], while keeping constant the component of the magnetic field perpendicular to the graphene plane ($B_\perp$), thus increasing the total magnetic field ($B_T$) applied to the sample. The effects related to the spin can be decoupled from those dependent on $B_\perp$ alone. For device A, the experiments in a tilted magnetic field were performed in two different cooldowns, indicated  by solid symbols (first cooldown) and open symbols (second cooldown) in Figs.~\ref{fig:fig3} and \ref{fig:fig4}.

In Fig.~\ref{fig:fig3}, we illustrate the splitting of the higher
LLs ($N> 0$) for different tilt angles. The minima in $R_{xx}$ and $C$ associated with even $\nu$ become more pronounced upon tilting the sample and increasing $B_T$, while those associated with the odd filling factors do not change. In particular, for $B_\perp = 8.4$~T and $\theta=0^\circ$, we observe only the standard sequence for the half-integer quantum Hall effect in graphene [Fig.~\ref{fig:fig3}(a) 
black line] while at $\theta= 73.7 ^\circ$ clear minima in $R_{xx}$ appear at the intermediate filling factors $\nu=-4, -8$, and $-12$ (red line). In the capacitance signal shown in Fig.\ref{fig:fig3}(c), the minimum at $\nu=4$ becomes progressively deeper as the sample is tilted  from 0$^\circ$ to 63.2$^\circ$, indicating that the DOS is reduced by an increase in $B_T$. In contrast, the minima at $\nu=-3$ and $-7$ in $R_{xx}$ [Fig.\ref{fig:fig3}(b)] and $\nu=3$ and $5$ in $C$ [Fig.\ref{fig:fig3}(c)] do not show any significant change as the sample is tilted.

We address quantitatively the splitting mechanism for the higher energy LLs extracting the activation gaps in a tilted magnetic field.
As Fig. \ref{fig:fig3}(d) shows, $\Delta_{-4}$, $\Delta_{-8}$, and $\Delta_{-12}$ increase linearly with  $B_{T}$ at a fixed $B_{\perp}= 10.4$~T and the gap size is smaller than the Zeeman energy (gray dashed line). The linear dependence of $\Delta_{-4}$ and $\Delta_{-8}$ on $B_T$ is found at several $B_\perp$ (10.4, 11.3, 20, and 25~T, the last two only for $\Delta_{-4}$).
At even $\nu$, the gap size can be described as a result of two separate contributions reduced by the Landau level broadening, $\Delta_{\nu}= E(B_\perp)+E_Z(B_T)-\Gamma$. The first term, $E(B_\perp)$, incorporates all the effects which depend only on $B_\perp$, (e.g., electron-electron interactions) and therefore does not change upon an increase of $B_T$. The value of $g$ in a titled field can be calculated by the derivative of $\Delta_{\nu}$ with respect to $B_T$, and it provides information about the spin of the excitation involved in the transport process \cite{SchmellerPRL1995}. The enhancement of $g$ due to exchange interactions depends solely upon $B_\perp$ and therefore does not influence the calculation. A linear fit to the data leads to $g\approx$~2 for the three filling factors at each $B_\perp$ meaning that transport takes place via thermally excited electron-hole pairs with reversed spin with no collective effects, such as skyrmions \cite{YangPRB2006}, involved. To illustrate the behavior of $g$, we plot $g$ as a function of $\nu$ at $B_\perp=10.4$ and 11.3~T as representative results in Fig.~\ref{fig:fig3}(e). 

In contrast to the even filling factors, the value of $\Delta_{-3}$ [Fig.\ref{fig:fig3}(f)] does not depend on $B_T$ and it is much smaller than the Zeeman energy. Therefore, we can assume that the origin of $\nu=-3$ lies in the lifting of the valley degeneracy in the $N=-1$ level.

We now turn our attention to $N=$~0, which, in a tilted magnetic field, behaves substantially different compared to the higher LLs. We first consider device A.  Figure \ref{fig:fig4}(a) shows that at $\nu=-1$ the minimum in $R_{xx}$ becomes deeper, tilting the sample from $\theta=0^\circ$ (black solid line) to $\theta=59.1^\circ$ (red solid line) at $B_\perp=14.1$ T. Accordingly, the gap associated with $\nu=-1$ increases with $B_T$ [Fig.\ref{fig:fig4}(b)]. The size of $\Delta_{-1}$ is larger than the Zeeman energy for $B_\perp\geq 20$~T, while it is compatible with $E_Z$ within the error bars at smaller $B_\perp$. If a finite Landau level width is taken into account, then one can see that an enhancement of $\Delta_{-1}$ with respect to $E_Z$ occurs also in the case of $B_\perp\leq 11.3$~T. Furthermore, as in the case of the even $\nu$ in $|N|>0$, we do not find any significant enhancement of $g$ from the dependence of $\Delta_{-1}$ on $B_T$ [see the dotted line in Fig.~\ref{fig:fig4}(b)]. This indicates that $\nu=-1$ separates two states with reversed spin and the enhancement of the gap size, with respect to $E_Z$, is governed by the exchange interaction due to $B_\perp$.

At half filling of $N=0$, the resistance maximum at $\nu=0$ decreases with the in-plane magnetic field [see Fig.\ref{fig:fig4}(c)] having $B_{\perp}\geq$2 T, confirming earlier observations on suspended \cite{ZhaoPRL2012} and h-BN supported samples \cite{YoungNatPhys2012}. This observation suggests that $\Delta_0$ is reduced upon increasing $B_T$, ruling out the scenario of a fully spin polarized state at half filled $N=0$.
 
The capacitance measurements on device B in tilted magnetic fields support the picture emerging from the transport experiments. Figure~\ref{fig:fig4}(d) shows that all three $\nu$ resulting from the splitting of the zero energy Landau level react to a change in $\theta$ at a fixed $B_\perp=10.1$~T. We also notice that the odd filling factors $\pm 1$ are enhanced by an increase in $B_T$, i.e., the gap size increases, while the minimum at $\nu=0$ becomes shallower as the sample is tilted from $\theta=0^\circ$ (black  solid line) and $\theta=59.7^\circ$ (red solid line), i.e., the gap decreases.

The observation of the full splitting of the LLs in a perpendicular field and in particular, the presence of odd filling factors in $|N|\neq0$, means that both samples show phenomenology typical of the quantum Hall ferromagnetism of graphene \cite{AliceaPRB2006,NomuraPRL2006}.
In agreement with the QHF picture, we observe enhanced energy gaps in perpendicular field for both $\nu=-1$ and $\nu=-4$. In addition, the square root dependence of $\Delta_{-3}$ on $B_\perp$ suggests that its origin is due to electron-electron interactions.

The titled field experiments enable us to compose a splitting scenario which is different for the $N=0$ level and the higher energy levels. In the lowest level, we find that the spin and the valley degeneracy are lifted at $\nu=\pm 1$ and $\nu=0$, respectively. Conversely, we find that for $N\neq 0$ the spin degeneracy is lifted at half filling while the valley degeneracy is lifted at quarter filling. It is worth pointing out the difference between $\nu=-1$ and the filling factors originating from the spin splitting in the higher levels: The size of $\Delta_{-1}$ cannot be explained solely by the Zeeman energy, but it must find an origin in the electron-electron interactions, whereas the gaps at $\nu=-4, -8$ and $-12$ appear to be Zeeman dominated. 

In conclusion, we have probed the lifting of the Landau level degeneracy of graphene with two different measurement techniques and we have measured the activated transport gap for most of the observed filling factors. Our experiments in tilted fields highlight the differences between $N=0$ and the higher energy levels, and we could probe the different splitting hierarchies in $N=\pm 1$ and $N=0$, confirming the findings of Ref.\onlinecite{YoungNatPhys2012}. However, probably due to the disorder in our sample, we did not find any indication for skyrmion excitations in the transport properties.

This work is part of the research programme of the 'Stichting voor Fundamenteel Onderzoek der Materie (FOM)', which is financially supported by the 'Nederlandse Organisatie voor Wetenschappelijk Onderzoek (NWO)'.
 
\bibliography{Bibliography}

\begin{thebibliography}{26}%
\makeatletter
\providecommand \@ifxundefined [1]{%
 \@ifx{#1\undefined}
}%
\providecommand \@ifnum [1]{%
 \ifnum #1\expandafter \@firstoftwo
 \else \expandafter \@secondoftwo
 \fi
}%
\providecommand \@ifx [1]{%
 \ifx #1\expandafter \@firstoftwo
 \else \expandafter \@secondoftwo
 \fi
}%
\providecommand \natexlab [1]{#1}%
\providecommand \enquote  [1]{``#1''}%
\providecommand \bibnamefont  [1]{#1}%
\providecommand \bibfnamefont [1]{#1}%
\providecommand \citenamefont [1]{#1}%
\providecommand \href@noop [0]{\@secondoftwo}%
\providecommand \href [0]{\begingroup \@sanitize@url \@href}%
\providecommand \@href[1]{\@@startlink{#1}\@@href}%
\providecommand \@@href[1]{\endgroup#1\@@endlink}%
\providecommand \@sanitize@url [0]{\catcode `\\12\catcode `\$12\catcode
  `\&12\catcode `\#12\catcode `\^12\catcode `\_12\catcode `\%12\relax}%
\providecommand \@@startlink[1]{}%
\providecommand \@@endlink[0]{}%
\providecommand \url  [0]{\begingroup\@sanitize@url \@url }%
\providecommand \@url [1]{\endgroup\@href {#1}{\urlprefix }}%
\providecommand \urlprefix  [0]{URL }%
\providecommand \Eprint [0]{\href }%
\providecommand \doibase [0]{http://dx.doi.org/}%
\providecommand \selectlanguage [0]{\@gobble}%
\providecommand \bibinfo  [0]{\@secondoftwo}%
\providecommand \bibfield  [0]{\@secondoftwo}%
\providecommand \translation [1]{[#1]}%
\providecommand \BibitemOpen [0]{}%
\providecommand \bibitemStop [0]{}%
\providecommand \bibitemNoStop [0]{.\EOS\space}%
\providecommand \EOS [0]{\spacefactor3000\relax}%
\providecommand \BibitemShut  [1]{\csname bibitem#1\endcsname}%
\let\auto@bib@innerbib\@empty
\bibitem [{\citenamefont {Novoselov}\ \emph {et~al.}(2005)\citenamefont
  {Novoselov}, \citenamefont {Geim}, \citenamefont {Morozov}, \citenamefont
  {Jiang}, \citenamefont {Katsnelson}, \citenamefont {Grigorieva},
  \citenamefont {Dubonos},\ and\ \citenamefont {Firsov}}]{NovoselovNat2005}%
  \BibitemOpen
  \bibfield  {author} {\bibinfo {author} {\bibfnamefont {K.~S.}\ \bibnamefont
  {Novoselov}}, \bibinfo {author} {\bibfnamefont {A.~K.}\ \bibnamefont {Geim}},
  \bibinfo {author} {\bibfnamefont {S.~V.}\ \bibnamefont {Morozov}}, \bibinfo
  {author} {\bibfnamefont {D.}~\bibnamefont {Jiang}}, \bibinfo {author}
  {\bibfnamefont {M.~I.}\ \bibnamefont {Katsnelson}}, \bibinfo {author}
  {\bibfnamefont {I.~V.}\ \bibnamefont {Grigorieva}}, \bibinfo {author}
  {\bibfnamefont {S.~V.}\ \bibnamefont {Dubonos}}, \ and\ \bibinfo {author}
  {\bibfnamefont {A.~A.}\ \bibnamefont {Firsov}},\ }\href@noop {} {\bibfield
  {journal} {\bibinfo  {journal} {Nature}\ }\textbf {\bibinfo {volume} {438}},\
  \bibinfo {pages} {197} (\bibinfo {year} {2005})}\BibitemShut {NoStop}%
\bibitem [{\citenamefont {Zhang}\ \emph {et~al.}(2005)\citenamefont {Zhang},
  \citenamefont {Tan}, \citenamefont {Stormer},\ and\ \citenamefont
  {Kim}}]{ZhangNat2005}%
  \BibitemOpen
  \bibfield  {author} {\bibinfo {author} {\bibfnamefont {Y.}~\bibnamefont
  {Zhang}}, \bibinfo {author} {\bibfnamefont {Y.-W.}\ \bibnamefont {Tan}},
  \bibinfo {author} {\bibfnamefont {H.~L.}\ \bibnamefont {Stormer}}, \ and\
  \bibinfo {author} {\bibfnamefont {P.}~\bibnamefont {Kim}},\ }\href {\doibase
  http://www.nature.com/nature/journal/v438/n7065/suppinfo/nature04235{\_}S1.html}
  {\bibfield  {journal} {\bibinfo  {journal} {Nature}\ }\textbf {\bibinfo
  {volume} {438}},\ \bibinfo {pages} {201} (\bibinfo {year}
  {2005})}\BibitemShut {NoStop}%
\bibitem [{\citenamefont {Alicea}\ and\ \citenamefont
  {Fisher}(2006)}]{AliceaPRB2006}%
  \BibitemOpen
  \bibfield  {author} {\bibinfo {author} {\bibfnamefont {J.}~\bibnamefont
  {Alicea}}\ and\ \bibinfo {author} {\bibfnamefont {M.~P.~A.}\ \bibnamefont
  {Fisher}},\ }\href {http://link.aps.org/doi/10.1103/PhysRevB.74.075422}
  {\bibfield  {journal} {\bibinfo  {journal} {Physical Review B}\ }\textbf
  {\bibinfo {volume} {74}},\ \bibinfo {pages} {75422} (\bibinfo {year}
  {2006})}\BibitemShut {NoStop}%
\bibitem [{\citenamefont {Sheng}\ \emph {et~al.}(2007)\citenamefont {Sheng},
  \citenamefont {Sheng}, \citenamefont {Haldane},\ and\ \citenamefont
  {Balents}}]{ShengPRL2007}%
  \BibitemOpen
  \bibfield  {author} {\bibinfo {author} {\bibfnamefont {L.}~\bibnamefont
  {Sheng}}, \bibinfo {author} {\bibfnamefont {D.~N.}\ \bibnamefont {Sheng}},
  \bibinfo {author} {\bibfnamefont {F.~D.~M.}\ \bibnamefont {Haldane}}, \ and\
  \bibinfo {author} {\bibfnamefont {L.}~\bibnamefont {Balents}},\ }\href
  {http://link.aps.org/doi/10.1103/PhysRevLett.99.196802} {\bibfield  {journal}
  {\bibinfo  {journal} {Physical Review Letters}\ }\textbf {\bibinfo {volume}
  {99}},\ \bibinfo {pages} {196802} (\bibinfo {year} {2007})}\BibitemShut
  {NoStop}%
\bibitem [{\citenamefont {Goerbig}(2011)}]{GoerbigRevModPhys2011}%
  \BibitemOpen
  \bibfield  {author} {\bibinfo {author} {\bibfnamefont {M.~O.}\ \bibnamefont
  {Goerbig}},\ }\href {http://link.aps.org/doi/10.1103/RevModPhys.83.1193}
  {\bibfield  {journal} {\bibinfo  {journal} {Reviews of Modern Physics}\
  }\textbf {\bibinfo {volume} {83}},\ \bibinfo {pages} {1193} (\bibinfo {year}
  {2011})}\BibitemShut {NoStop}%
\bibitem [{\citenamefont {Herbut}(2007)}]{HerbutPRB2007}%
  \BibitemOpen
  \bibfield  {author} {\bibinfo {author} {\bibfnamefont {I.~F.}\ \bibnamefont
  {Herbut}},\ }\href {http://link.aps.org/doi/10.1103/PhysRevB.75.165411}
  {\bibfield  {journal} {\bibinfo  {journal} {Physical Review B}\ }\textbf
  {\bibinfo {volume} {75}},\ \bibinfo {pages} {165411} (\bibinfo {year}
  {2007})}\BibitemShut {NoStop}%
\bibitem [{\citenamefont {Gusynin}\ \emph {et~al.}(2006)\citenamefont
  {Gusynin}, \citenamefont {Miransky}, \citenamefont {Sharapov},\ and\
  \citenamefont {Shovkovy}}]{GusyninPRB2006}%
  \BibitemOpen
  \bibfield  {author} {\bibinfo {author} {\bibfnamefont {V.~P.}\ \bibnamefont
  {Gusynin}}, \bibinfo {author} {\bibfnamefont {V.~A.}\ \bibnamefont
  {Miransky}}, \bibinfo {author} {\bibfnamefont {S.~G.}\ \bibnamefont
  {Sharapov}}, \ and\ \bibinfo {author} {\bibfnamefont {I.~A.}\ \bibnamefont
  {Shovkovy}},\ }\href {http://link.aps.org/doi/10.1103/PhysRevB.74.195429}
  {\bibfield  {journal} {\bibinfo  {journal} {Physical Review B}\ }\textbf
  {\bibinfo {volume} {74}},\ \bibinfo {pages} {195429} (\bibinfo {year}
  {2006})}\BibitemShut {NoStop}%
\bibitem [{\citenamefont {Luo}\ and\ \citenamefont
  {C{\^{o}}t{\'{e}}}(2013)}]{LuoPRB2013}%
  \BibitemOpen
  \bibfield  {author} {\bibinfo {author} {\bibfnamefont {W.}~\bibnamefont
  {Luo}}\ and\ \bibinfo {author} {\bibfnamefont {R.}~\bibnamefont
  {C{\^{o}}t{\'{e}}}},\ }\href
  {http://link.aps.org/doi/10.1103/PhysRevB.88.115417} {\bibfield  {journal}
  {\bibinfo  {journal} {Physical Review B}\ }\textbf {\bibinfo {volume} {88}},\
  \bibinfo {pages} {115417} (\bibinfo {year} {2013})}\BibitemShut {NoStop}%
\bibitem [{\citenamefont {Roy}(2011)}]{RoyPRB2011}%
  \BibitemOpen
  \bibfield  {author} {\bibinfo {author} {\bibfnamefont {B.}~\bibnamefont
  {Roy}},\ }\href {http://link.aps.org/doi/10.1103/PhysRevB.84.035458}
  {\bibfield  {journal} {\bibinfo  {journal} {Physical Review B}\ }\textbf
  {\bibinfo {volume} {84}},\ \bibinfo {pages} {35458} (\bibinfo {year}
  {2011})}\BibitemShut {NoStop}%
\bibitem [{\citenamefont {Zhang}\ \emph {et~al.}(2006)\citenamefont {Zhang},
  \citenamefont {Jiang}, \citenamefont {Small}, \citenamefont {Purewal},
  \citenamefont {Tan}, \citenamefont {Fazlollahi}, \citenamefont {Chudow},
  \citenamefont {Jaszczak}, \citenamefont {Stormer},\ and\ \citenamefont
  {Kim}}]{ZhangPRL2006}%
  \BibitemOpen
  \bibfield  {author} {\bibinfo {author} {\bibfnamefont {Y.}~\bibnamefont
  {Zhang}}, \bibinfo {author} {\bibfnamefont {Z.}~\bibnamefont {Jiang}},
  \bibinfo {author} {\bibfnamefont {J.~P.}\ \bibnamefont {Small}}, \bibinfo
  {author} {\bibfnamefont {M.~S.}\ \bibnamefont {Purewal}}, \bibinfo {author}
  {\bibfnamefont {Y.~W.}\ \bibnamefont {Tan}}, \bibinfo {author} {\bibfnamefont
  {M.}~\bibnamefont {Fazlollahi}}, \bibinfo {author} {\bibfnamefont {J.~D.}\
  \bibnamefont {Chudow}}, \bibinfo {author} {\bibfnamefont {J.~A.}\
  \bibnamefont {Jaszczak}}, \bibinfo {author} {\bibfnamefont {H.~L.}\
  \bibnamefont {Stormer}}, \ and\ \bibinfo {author} {\bibfnamefont
  {P.}~\bibnamefont {Kim}},\ }\href
  {http://link.aps.org/doi/10.1103/PhysRevLett.96.136806} {\bibfield  {journal}
  {\bibinfo  {journal} {Physical Review Letters}\ }\textbf {\bibinfo {volume}
  {96}},\ \bibinfo {pages} {136806} (\bibinfo {year} {2006})}\BibitemShut
  {NoStop}%
\bibitem [{\citenamefont {Jiang}\ \emph {et~al.}(2007)\citenamefont {Jiang},
  \citenamefont {Zhang}, \citenamefont {Stormer},\ and\ \citenamefont
  {Kim}}]{JiangPRL2007}%
  \BibitemOpen
  \bibfield  {author} {\bibinfo {author} {\bibfnamefont {Z.}~\bibnamefont
  {Jiang}}, \bibinfo {author} {\bibfnamefont {Y.}~\bibnamefont {Zhang}},
  \bibinfo {author} {\bibfnamefont {H.~L.}\ \bibnamefont {Stormer}}, \ and\
  \bibinfo {author} {\bibfnamefont {P.}~\bibnamefont {Kim}},\ }\href
  {http://link.aps.org/doi/10.1103/PhysRevLett.99.106802} {\bibfield  {journal}
  {\bibinfo  {journal} {Physical Review Letters}\ }\textbf {\bibinfo {volume}
  {99}},\ \bibinfo {pages} {106802} (\bibinfo {year} {2007})}\BibitemShut
  {NoStop}%
\bibitem [{\citenamefont {Zhang}\ \emph {et~al.}(2009)\citenamefont {Zhang},
  \citenamefont {Camacho}, \citenamefont {Cao}, \citenamefont {Chen},
  \citenamefont {Khodas}, \citenamefont {Kharzeev}, \citenamefont {Tsvelik},
  \citenamefont {Valla},\ and\ \citenamefont {Zaliznyak}}]{ZhangPRB2009}%
  \BibitemOpen
  \bibfield  {author} {\bibinfo {author} {\bibfnamefont {L.}~\bibnamefont
  {Zhang}}, \bibinfo {author} {\bibfnamefont {J.}~\bibnamefont {Camacho}},
  \bibinfo {author} {\bibfnamefont {H.}~\bibnamefont {Cao}}, \bibinfo {author}
  {\bibfnamefont {Y.~P.}\ \bibnamefont {Chen}}, \bibinfo {author}
  {\bibfnamefont {M.}~\bibnamefont {Khodas}}, \bibinfo {author} {\bibfnamefont
  {D.~E.}\ \bibnamefont {Kharzeev}}, \bibinfo {author} {\bibfnamefont {A.~M.}\
  \bibnamefont {Tsvelik}}, \bibinfo {author} {\bibfnamefont {T.}~\bibnamefont
  {Valla}}, \ and\ \bibinfo {author} {\bibfnamefont {I.~A.}\ \bibnamefont
  {Zaliznyak}},\ }\href {http://link.aps.org/doi/10.1103/PhysRevB.80.241412}
  {\bibfield  {journal} {\bibinfo  {journal} {Physical Review B}\ }\textbf
  {\bibinfo {volume} {80}},\ \bibinfo {pages} {241412} (\bibinfo {year}
  {2009})}\BibitemShut {NoStop}%
\bibitem [{\citenamefont {Giesbers}\ \emph {et~al.}(2009)\citenamefont
  {Giesbers}, \citenamefont {Ponomarenko}, \citenamefont {Novoselov},
  \citenamefont {Geim}, \citenamefont {Katsnelson}, \citenamefont {Maan},\ and\
  \citenamefont {Zeitler}}]{GiesbersPRB2009}%
  \BibitemOpen
  \bibfield  {author} {\bibinfo {author} {\bibfnamefont {A.~J.~M.}\
  \bibnamefont {Giesbers}}, \bibinfo {author} {\bibfnamefont {L.~A.}\
  \bibnamefont {Ponomarenko}}, \bibinfo {author} {\bibfnamefont {K.~S.}\
  \bibnamefont {Novoselov}}, \bibinfo {author} {\bibfnamefont {A.~K.}\
  \bibnamefont {Geim}}, \bibinfo {author} {\bibfnamefont {M.~I.}\ \bibnamefont
  {Katsnelson}}, \bibinfo {author} {\bibfnamefont {J.~C.}\ \bibnamefont
  {Maan}}, \ and\ \bibinfo {author} {\bibfnamefont {U.}~\bibnamefont
  {Zeitler}},\ }\href {http://link.aps.org/doi/10.1103/PhysRevB.80.201403}
  {\bibfield  {journal} {\bibinfo  {journal} {Physical Review B}\ }\textbf
  {\bibinfo {volume} {80}},\ \bibinfo {pages} {201403} (\bibinfo {year}
  {2009})}\BibitemShut {NoStop}%
\bibitem [{\citenamefont {Checkelsky}\ \emph {et~al.}(2008)\citenamefont
  {Checkelsky}, \citenamefont {Li},\ and\ \citenamefont
  {Ong}}]{CheckelskyPRL2008}%
  \BibitemOpen
  \bibfield  {author} {\bibinfo {author} {\bibfnamefont {J.~G.}\ \bibnamefont
  {Checkelsky}}, \bibinfo {author} {\bibfnamefont {L.}~\bibnamefont {Li}}, \
  and\ \bibinfo {author} {\bibfnamefont {N.~P.}\ \bibnamefont {Ong}},\ }\href
  {http://link.aps.org/doi/10.1103/PhysRevLett.100.206801} {\bibfield
  {journal} {\bibinfo  {journal} {Physical Review Letters}\ }\textbf {\bibinfo
  {volume} {100}},\ \bibinfo {pages} {206801} (\bibinfo {year}
  {2008})}\BibitemShut {NoStop}%
\bibitem [{\citenamefont {Young}\ \emph {et~al.}(2012)\citenamefont {Young},
  \citenamefont {Dean}, \citenamefont {Wang}, \citenamefont {Ren},
  \citenamefont {Cadden-Zimansky}, \citenamefont {Watanabe}, \citenamefont
  {Taniguchi}, \citenamefont {Hone}, \citenamefont {Shepard},\ and\
  \citenamefont {Kim}}]{YoungNatPhys2012}%
  \BibitemOpen
  \bibfield  {author} {\bibinfo {author} {\bibfnamefont {A.~F.}\ \bibnamefont
  {Young}}, \bibinfo {author} {\bibfnamefont {C.~R.}\ \bibnamefont {Dean}},
  \bibinfo {author} {\bibfnamefont {L.}~\bibnamefont {Wang}}, \bibinfo {author}
  {\bibfnamefont {H.}~\bibnamefont {Ren}}, \bibinfo {author} {\bibfnamefont
  {P.}~\bibnamefont {Cadden-Zimansky}}, \bibinfo {author} {\bibfnamefont
  {K.}~\bibnamefont {Watanabe}}, \bibinfo {author} {\bibfnamefont
  {T.}~\bibnamefont {Taniguchi}}, \bibinfo {author} {\bibfnamefont
  {J.}~\bibnamefont {Hone}}, \bibinfo {author} {\bibfnamefont {K.~L.}\
  \bibnamefont {Shepard}}, \ and\ \bibinfo {author} {\bibfnamefont
  {P.}~\bibnamefont {Kim}},\ }\href {\doibase
  http://www.nature.com/nphys/journal/v8/n7/abs/nphys2307.html{\#}supplementary-information}
  {\bibfield  {journal} {\bibinfo  {journal} {Nat Phys}\ }\textbf {\bibinfo
  {volume} {8}},\ \bibinfo {pages} {550} (\bibinfo {year} {2012})}\BibitemShut
  {NoStop}%
\bibitem [{\citenamefont {Zhao}\ \emph {et~al.}(2012)\citenamefont {Zhao},
  \citenamefont {Cadden-Zimansky}, \citenamefont {Ghahari},\ and\ \citenamefont
  {Kim}}]{ZhaoPRL2012}%
  \BibitemOpen
  \bibfield  {author} {\bibinfo {author} {\bibfnamefont {Y.}~\bibnamefont
  {Zhao}}, \bibinfo {author} {\bibfnamefont {P.}~\bibnamefont
  {Cadden-Zimansky}}, \bibinfo {author} {\bibfnamefont {F.}~\bibnamefont
  {Ghahari}}, \ and\ \bibinfo {author} {\bibfnamefont {P.}~\bibnamefont
  {Kim}},\ }\href {http://link.aps.org/doi/10.1103/PhysRevLett.108.106804}
  {\bibfield  {journal} {\bibinfo  {journal} {Physical Review Letters}\
  }\textbf {\bibinfo {volume} {108}},\ \bibinfo {pages} {106804} (\bibinfo
  {year} {2012})}\BibitemShut {NoStop}%
\bibitem [{\citenamefont {Amet}\ \emph {et~al.}(2014)\citenamefont {Amet},
  \citenamefont {Williams}, \citenamefont {Watanabe}, \citenamefont
  {Taniguchi},\ and\ \citenamefont {Goldhaber-Gordon}}]{AmetPRL2014}%
  \BibitemOpen
  \bibfield  {author} {\bibinfo {author} {\bibfnamefont {F.}~\bibnamefont
  {Amet}}, \bibinfo {author} {\bibfnamefont {J.~R.}\ \bibnamefont {Williams}},
  \bibinfo {author} {\bibfnamefont {K.}~\bibnamefont {Watanabe}}, \bibinfo
  {author} {\bibfnamefont {T.}~\bibnamefont {Taniguchi}}, \ and\ \bibinfo
  {author} {\bibfnamefont {D.}~\bibnamefont {Goldhaber-Gordon}},\ }\href
  {http://link.aps.org/doi/10.1103/PhysRevLett.112.196601} {\bibfield
  {journal} {\bibinfo  {journal} {Physical Review Letters}\ }\textbf {\bibinfo
  {volume} {112}},\ \bibinfo {pages} {196601} (\bibinfo {year}
  {2014})}\BibitemShut {NoStop}%
\bibitem [{\citenamefont {Yu}\ \emph {et~al.}(2013)\citenamefont {Yu},
  \citenamefont {Jalil}, \citenamefont {Belle}, \citenamefont {Mayorov},
  \citenamefont {Blake}, \citenamefont {Schedin}, \citenamefont {Morozov},
  \citenamefont {Ponomarenko}, \citenamefont {Chiappini}, \citenamefont
  {Wiedmann}, \citenamefont {Zeitler}, \citenamefont {Katsnelson},
  \citenamefont {Geim}, \citenamefont {Novoselov},\ and\ \citenamefont
  {Elias}}]{YuPNAS2013}%
  \BibitemOpen
  \bibfield  {author} {\bibinfo {author} {\bibfnamefont {G.~L.}\ \bibnamefont
  {Yu}}, \bibinfo {author} {\bibfnamefont {R.}~\bibnamefont {Jalil}}, \bibinfo
  {author} {\bibfnamefont {B.}~\bibnamefont {Belle}}, \bibinfo {author}
  {\bibfnamefont {A.~S.}\ \bibnamefont {Mayorov}}, \bibinfo {author}
  {\bibfnamefont {P.}~\bibnamefont {Blake}}, \bibinfo {author} {\bibfnamefont
  {F.}~\bibnamefont {Schedin}}, \bibinfo {author} {\bibfnamefont {S.~V.}\
  \bibnamefont {Morozov}}, \bibinfo {author} {\bibfnamefont {L.~A.}\
  \bibnamefont {Ponomarenko}}, \bibinfo {author} {\bibfnamefont
  {F.}~\bibnamefont {Chiappini}}, \bibinfo {author} {\bibfnamefont
  {S.}~\bibnamefont {Wiedmann}}, \bibinfo {author} {\bibfnamefont
  {U.}~\bibnamefont {Zeitler}}, \bibinfo {author} {\bibfnamefont {M.~I.}\
  \bibnamefont {Katsnelson}}, \bibinfo {author} {\bibfnamefont {A.~K.}\
  \bibnamefont {Geim}}, \bibinfo {author} {\bibfnamefont {K.~S.}\ \bibnamefont
  {Novoselov}}, \ and\ \bibinfo {author} {\bibfnamefont {D.~C.}\ \bibnamefont
  {Elias}},\ }\href {http://www.pnas.org/content/110/9/3282.abstract}
  {\bibfield  {journal} {\bibinfo  {journal} {Proceedings of the National
  Academy of Sciences}\ }\textbf {\bibinfo {volume} {110}},\ \bibinfo {pages}
  {3282} (\bibinfo {year} {2013})}\BibitemShut {NoStop}%
\bibitem [{\citenamefont {Nomura}\ and\ \citenamefont
  {MacDonald}(2006)}]{NomuraPRL2006}%
  \BibitemOpen
  \bibfield  {author} {\bibinfo {author} {\bibfnamefont {K.}~\bibnamefont
  {Nomura}}\ and\ \bibinfo {author} {\bibfnamefont {A.~H.}\ \bibnamefont
  {MacDonald}},\ }\href {http://link.aps.org/doi/10.1103/PhysRevLett.96.256602}
  {\bibfield  {journal} {\bibinfo  {journal} {Physical Review Letters}\
  }\textbf {\bibinfo {volume} {96}},\ \bibinfo {pages} {256602} (\bibinfo
  {year} {2006})}\BibitemShut {NoStop}%
\bibitem [{\citenamefont {Young}\ \emph {et~al.}(2013)\citenamefont {Young},
  \citenamefont {Sanchez-Yamagishi}, \citenamefont {Hunt}, \citenamefont
  {Choi}, \citenamefont {Watanabe}, \citenamefont {Taniguchi}, \citenamefont
  {Ashoori},\ and\ \citenamefont {Jarillo-Herrero}}]{YoungNature2013}%
  \BibitemOpen
  \bibfield  {author} {\bibinfo {author} {\bibfnamefont {A.~F.}\ \bibnamefont
  {Young}}, \bibinfo {author} {\bibfnamefont {J.~D.}\ \bibnamefont
  {Sanchez-Yamagishi}}, \bibinfo {author} {\bibfnamefont {B.}~\bibnamefont
  {Hunt}}, \bibinfo {author} {\bibfnamefont {S.~H.}\ \bibnamefont {Choi}},
  \bibinfo {author} {\bibfnamefont {K.}~\bibnamefont {Watanabe}}, \bibinfo
  {author} {\bibfnamefont {T.}~\bibnamefont {Taniguchi}}, \bibinfo {author}
  {\bibfnamefont {R.~C.}\ \bibnamefont {Ashoori}}, \ and\ \bibinfo {author}
  {\bibfnamefont {P.}~\bibnamefont {Jarillo-Herrero}},\ }\href {\doibase
  10.1038/nature12800} {\bibfield  {journal} {\bibinfo  {journal} {Nature}\
  }\textbf {\bibinfo {volume} {505}},\ \bibinfo {pages} {528} (\bibinfo {year}
  {2013})}\BibitemShut {NoStop}%
\bibitem [{\citenamefont {Luryi}(1988)}]{LuryiApplPhysLett1988}%
  \BibitemOpen
  \bibfield  {author} {\bibinfo {author} {\bibfnamefont {S.}~\bibnamefont
  {Luryi}},\ }\href {\doibase doi:http://dx.doi.org/10.1063/1.99649} {\bibfield
   {journal} {\bibinfo  {journal} {Applied Physics Letters}\ }\textbf {\bibinfo
  {volume} {52}},\ \bibinfo {pages} {501} (\bibinfo {year} {1988})}\BibitemShut
  {NoStop}%
\bibitem [{\citenamefont {Dean}\ \emph {et~al.}(2010)\citenamefont {Dean},
  \citenamefont {Young}, \citenamefont {Meric}, \citenamefont {Lee},
  \citenamefont {Wang}, \citenamefont {Sorgenfrei}, \citenamefont {Watanabe},
  \citenamefont {Taniguchi}, \citenamefont {Kim}, \citenamefont {Shepard},\
  and\ \citenamefont {Hone}}]{DeanNatNanotech2010}%
  \BibitemOpen
  \bibfield  {author} {\bibinfo {author} {\bibfnamefont {C.~R.}\ \bibnamefont
  {Dean}}, \bibinfo {author} {\bibfnamefont {A.~F.}\ \bibnamefont {Young}},
  \bibinfo {author} {\bibfnamefont {I.}~\bibnamefont {Meric}}, \bibinfo
  {author} {\bibfnamefont {C.}~\bibnamefont {Lee}}, \bibinfo {author}
  {\bibfnamefont {L.}~\bibnamefont {Wang}}, \bibinfo {author} {\bibfnamefont
  {S.}~\bibnamefont {Sorgenfrei}}, \bibinfo {author} {\bibfnamefont
  {K.}~\bibnamefont {Watanabe}}, \bibinfo {author} {\bibfnamefont
  {T.}~\bibnamefont {Taniguchi}}, \bibinfo {author} {\bibfnamefont
  {P.}~\bibnamefont {Kim}}, \bibinfo {author} {\bibfnamefont {K.~L.}\
  \bibnamefont {Shepard}}, \ and\ \bibinfo {author} {\bibfnamefont
  {J.}~\bibnamefont {Hone}},\ }\href {\doibase
  http://www.nature.com/nnano/journal/v5/n10/abs/nnano.2010.172.html{\#}supplementary-information}
  {\bibfield  {journal} {\bibinfo  {journal} {Nat Nano}\ }\textbf {\bibinfo
  {volume} {5}},\ \bibinfo {pages} {722} (\bibinfo {year} {2010})}\BibitemShut
  {NoStop}%
\bibitem [{\citenamefont {Kurganova}\ \emph {et~al.}(2010)\citenamefont
  {Kurganova}, \citenamefont {Giesbers}, \citenamefont {Gorbachev},
  \citenamefont {Geim}, \citenamefont {Novoselov}, \citenamefont {Maan},\ and\
  \citenamefont {Zeitler}}]{KurganovaSolidStateComm_2010}%
  \BibitemOpen
  \bibfield  {author} {\bibinfo {author} {\bibfnamefont {E.~V.}\ \bibnamefont
  {Kurganova}}, \bibinfo {author} {\bibfnamefont {A.~J.~M.}\ \bibnamefont
  {Giesbers}}, \bibinfo {author} {\bibfnamefont {R.~V.}\ \bibnamefont
  {Gorbachev}}, \bibinfo {author} {\bibfnamefont {A.~K.}\ \bibnamefont {Geim}},
  \bibinfo {author} {\bibfnamefont {K.~S.}\ \bibnamefont {Novoselov}}, \bibinfo
  {author} {\bibfnamefont {J.~C.}\ \bibnamefont {Maan}}, \ and\ \bibinfo
  {author} {\bibfnamefont {U.}~\bibnamefont {Zeitler}},\ }\href {\doibase
  http://dx.doi.org/10.1016/j.ssc.2010.09.042} {\bibfield  {journal} {\bibinfo
  {journal} {Solid State Communications}\ }\textbf {\bibinfo {volume} {150}},\
  \bibinfo {pages} {2209} (\bibinfo {year} {2010})}\BibitemShut {NoStop}%
\bibitem [{\citenamefont {Kurganova}\ \emph {et~al.}(2011)\citenamefont
  {Kurganova}, \citenamefont {van Elferen}, \citenamefont {McCollam},
  \citenamefont {Ponomarenko}, \citenamefont {Novoselov}, \citenamefont
  {Veligura}, \citenamefont {van Wees}, \citenamefont {Maan},\ and\
  \citenamefont {Zeitler}}]{KurganovaPRB_2011}%
  \BibitemOpen
  \bibfield  {author} {\bibinfo {author} {\bibfnamefont {E.~V.}\ \bibnamefont
  {Kurganova}}, \bibinfo {author} {\bibfnamefont {H.~J.}\ \bibnamefont {van
  Elferen}}, \bibinfo {author} {\bibfnamefont {A.}~\bibnamefont {McCollam}},
  \bibinfo {author} {\bibfnamefont {L.~A.}\ \bibnamefont {Ponomarenko}},
  \bibinfo {author} {\bibfnamefont {K.~S.}\ \bibnamefont {Novoselov}}, \bibinfo
  {author} {\bibfnamefont {A.}~\bibnamefont {Veligura}}, \bibinfo {author}
  {\bibfnamefont {B.~J.}\ \bibnamefont {van Wees}}, \bibinfo {author}
  {\bibfnamefont {J.~C.}\ \bibnamefont {Maan}}, \ and\ \bibinfo {author}
  {\bibfnamefont {U.}~\bibnamefont {Zeitler}},\ }\href
  {http://link.aps.org/doi/10.1103/PhysRevB.84.121407} {\bibfield  {journal}
  {\bibinfo  {journal} {Physical Review B}\ }\textbf {\bibinfo {volume} {84}},\
  \bibinfo {pages} {121407} (\bibinfo {year} {2011})}\BibitemShut {NoStop}%
\bibitem [{\citenamefont {Schmeller}\ \emph {et~al.}(1995)\citenamefont
  {Schmeller}, \citenamefont {Eisenstein}, \citenamefont {Pfeiffer},\ and\
  \citenamefont {West}}]{SchmellerPRL1995}%
  \BibitemOpen
  \bibfield  {author} {\bibinfo {author} {\bibfnamefont {A.}~\bibnamefont
  {Schmeller}}, \bibinfo {author} {\bibfnamefont {J.~P.}\ \bibnamefont
  {Eisenstein}}, \bibinfo {author} {\bibfnamefont {L.~N.}\ \bibnamefont
  {Pfeiffer}}, \ and\ \bibinfo {author} {\bibfnamefont {K.~W.}\ \bibnamefont
  {West}},\ }\href {\doibase 10.1103/PhysRevLett.75.4290} {\bibfield  {journal}
  {\bibinfo  {journal} {Physical Review Letters}\ }\textbf {\bibinfo {volume}
  {75}},\ \bibinfo {pages} {4290} (\bibinfo {year} {1995})}\BibitemShut
  {NoStop}%
\bibitem [{\citenamefont {Yang}\ \emph {et~al.}(2006)\citenamefont {Yang},
  \citenamefont {{Das Sarma}},\ and\ \citenamefont {MacDonald}}]{YangPRB2006}%
  \BibitemOpen
  \bibfield  {author} {\bibinfo {author} {\bibfnamefont {K.}~\bibnamefont
  {Yang}}, \bibinfo {author} {\bibfnamefont {S.}~\bibnamefont {{Das Sarma}}}, \
  and\ \bibinfo {author} {\bibfnamefont {A.~H.}\ \bibnamefont {MacDonald}},\
  }\href {http://link.aps.org/doi/10.1103/PhysRevB.74.075423} {\bibfield
  {journal} {\bibinfo  {journal} {Physical Review B}\ }\textbf {\bibinfo
  {volume} {74}},\ \bibinfo {pages} {75423} (\bibinfo {year}
  {2006})}\BibitemShut {NoStop}%
\end{thebibliography}%

\end{document}